\documentclass[aps,prb,twocolumn,superscriptaddress,letterpaper]{revtex4}
\usepackage{amsmath}
\usepackage{dcolumn}
\usepackage{epsfig}
\usepackage{graphicx}
\usepackage{latexsym}
\usepackage{amssymb}
\usepackage{color}
\usepackage{amsfonts}
\usepackage{bm}

\newcommand{\be}{\begin{eqnarray}}
\newcommand{\ee}{\end{eqnarray}}
\newcommand{\ba}{\begin{array}}
\newcommand{\ea}{\end{array}}

\newcommand{\sign}{\mathop{\rm sign}\nolimits}

\begin{document}
\title{Hall conductivity dominated by fluctuations near the superconducting transition in disordered films}

\author{Nicholas P. Breznay}
\affiliation{Department of Applied Physics, Stanford University, Stanford, CA 94305, USA}
\author{Karen Michaeli}
\affiliation{Department of Physics, Massachusetts Institute of Technology, Cambridge, MA 02139}
\author{Konstantin S. Tikhonov}
\affiliation{Department of Physics, Texas A$\&$M University, College Station, TX 77843}
\author{Alexander M. Finkel'stein}
\affiliation{Department of Physics, Texas A$\&$M University, College Station, TX 77843}
\affiliation{Department of Condensed Matter Physics, The Weizmann Institute of Science, Rehovot 76100, Israel}
\author{Mihir Tendulkar}
\affiliation{Department of Applied Physics, Stanford University, Stanford, CA 94305, USA}
\author{Aharon Kapitulnik}
\affiliation{Department of Applied Physics, Stanford University, Stanford, CA 94305, USA}
\affiliation{Department of Physics, Stanford University, Stanford, CA 94305, USA}

\date{\today}

\begin{abstract}

We have studied the Hall effect in superconducting tantalum nitride films.  We find a large contribution to the Hall conductivity near the superconducting transition, which we can track to temperatures well above $T_c$ and magnetic fields well above the upper critical field, $\text{\text{H}}_{c2}(0)$. This contribution arises from Aslamazov-Larkin superconducting fluctuations, and we find quantitative agreement between our data and recent theoretical analysis based on time dependent Ginzburg-Landau theory.

\end{abstract}

\maketitle

\section{Introduction}

Thin superconducting films are characterized by reduced dimensionality and short coherence length, both of which contribute to the enhancement of fluctuations of the superconducting order parameter above the transition temperature.  These fluctuations are expected to affect both thermodynamic and transport measurements.  Properties such as the specific heat, magnetization and electrical conductivity in the vicinity of the superconducting transition have been studied both experimentally~\cite{skoctink} and theoretically.~\cite{varlamov} In particular, fluctuation effects on the electrical conductivity were first discovered by Glover,\cite{glover} and the diagonal elements of the conductivity tensor (or paraconductivity) are now well understood following the original work of Aslamazov and Larkin \cite{aslam}, Maki \cite{maki1968} and Thompson .\cite{thomp1970}  \text{H}owever, comparable experimental studies of the off diagonal (\text{H}all) conductivity have been relatively limited in scope.

In this paper we investigate the longitudinal and \text{H}all conductivities of ultrathin disordered tantalum nitride (TaN$_x$) films as a function of perpendicular magnetic field close to and above the zero field critical temperature $T_{c0}=T_c(\text{\text{H}}=0)$.  Although both the longitudinal ($R_{xx}$) and \text{H}all ($R_{xy}$) resistances vanish in the superconducting state, we find an enhanced \text{H}all resistance above the superconducting transition temperature $T_c(\text{\text{H}})$.  Such an enhanced resistance can be understood by considering dominant contributions of time-dependent fluctuation effects to the full conductivity tensor above $T_c$.

The \text{H}all effect at temperatures near $T_c$ has been studied in thin films of conventional superconductors such as MoSi, MoGe, NbGe, and amorphous InO,~\cite{smith94, graybeal94,kokubo01,paalanen92} as well as in strongly anisotropic cuprate superconductors.~\cite{hagen93,samoilov94,lang9495,liu97} Nevertheless, it is not well understood and continues to be a topic of active research.~\cite{puica09}  For example, an unexpected sign reversal of the \text{H}all voltage near $T_c$ has sparked considerable debate (see e.g. Refs.~\onlinecite{hagen93} and~\onlinecite{samoilov94} and references therein).  \text{H}owever, all of these studies have been complicated by vortex physics  below $T_c$ or by contributions from the normal state \text{H}all effect.  A number of  microscopic and phenomenological studies  have considered contributions due to vortices, pinning effects,~\cite{dorsey,zhu99} and superconducting fluctuations.~\cite{fukuyama71,ullah91,aronov95}  Efforts to reconcile these studies have been hampered by the difficulty of probing fluctuation effects in the \text{H}all conductivity in conventional superconducting films.  Challenges include the combination of high carrier concentration and large longitudinal resistance typical for such systems.  Thus, no conclusive picture for the effect of fluctuations of  the superconducting order parameter on the \text{H}all conductivity in the normal phase has been reached.  In this paper we study field-dependent fluctuation effects in a regime where they may be unambiguously separated from vortex physics and from distinct normal state contributions. Thus, we provide a complete description of the \text{H}all effect in a thin disordered film close to the phase transition into the superconducting state.

\section{Experimental Background}

Tantalum nitride films were prepared using sputter deposition onto a Si substrate.  Sample composition was analyzed using x-ray photoemission spectroscopy and determined to be $50\pm10$ at. $\%$ $N$.  Sample thicknesses were well controlled using the sputtering time and confirmed via x-ray reflectivity and TEM measurements; the sample thickness $d$ is 4.9 nm. X-ray diffraction analysis showed no sign of crystalline order, and surface analyses showed no signs of granularity or inhomogeneity. Samples were patterned into \text{H}all bar devices using standard optical photolithography techniques and Ar-ion etching, and Ti-Au electrical contact pads were deposited using electron beam evaporation.  The active area of the devices is 400 $\mu$m $\times$ 100 $\mu$m. Linear longitudinal and \text{H}all resistance were measured using standard four-point low frequency lock-in techniques in perpendicular magnetic fields; care was taken to ensure that all measurements were linear in the excitation current.  The \text{H}all resistance was extracted from the component of the \text{H}all voltage antisymmetric in the applied field, and was typically $\sim$~100 times smaller than the longitudinal contribution. Five devices fabricated from the same film were measured and all demonstrated qualitatively identical behavior; the results presented in this paper are from two representative devices.  Throughout the paper the longitudinal and \text{H}all resistance and conductivity data and theoretical expressions are given in two dimensional (sheet) quantities.

	\begin{figure}
	\centering
	\includegraphics[width=\columnwidth]{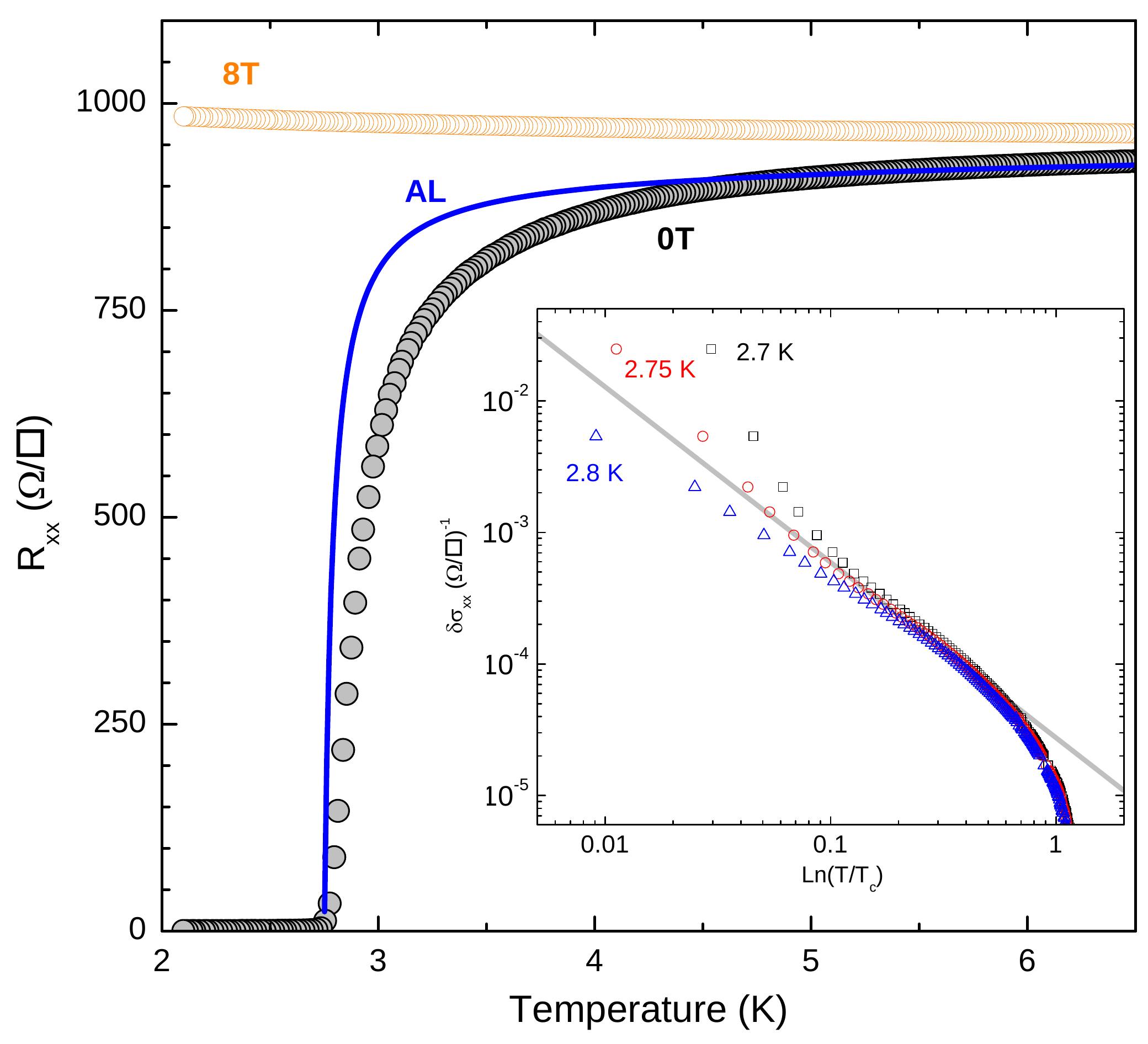}
	\caption{ \footnotesize \setlength{\baselineskip}{0.8\baselineskip} Resistance versus temperature of sample 1, a 4.9-nm-thick tantalum nitride film, in zero magnetic field and in a field of 8 T. The continuous curve shows the expected Aslamsov-Larkin enhancement in the conductivity above $T_{c0} \approx$ 2.75 K. The inset shows the measured fluctuation conductivity $\delta \sigma_{xx}$ for this sample plotted versus $ln(T/T_{c0})$ computed using three values of $T_{c0}$; the solid line depicts a slope of -1.33 which is expected for an AL term that is percolation dominated.}
	\label{fig:resistive}
	\end{figure}

\section{Longitudinal Resistance}

To understand the Hall effect, it is important to first understand the behavior of the longitudinal resistance.  Outside the superconducting phase and away from to the transition, the system is characterized by two types of low energy degrees of freedom: quasiparticles that are described using Fermi liquid theory and superconducting fluctuations.  The normal state conductivity $\sigma_{n}$ far from $T_c$ is attributed to the quasiparticles.  In the vicinity of the transition, the fluctuations of the superconducting order parameter create a new channel for  the electric current.  The main contribution of the superconducting fluctuations to the electrical transport can be formulated as the ``Drude term'' for these degrees of freedom.  This contribution corresponds to the Aslamazov-Larkin (AL) term.~\cite{varlamov,aslam}  To estimate the AL term one has to find the lifetime of the superconducting fluctuations, $\tau_{sc}$, because the Drude-like conductivity is proportional to it.  The finite lifetime of the superconducting fluctuations reflects the fact that outside of the superconducting phase the creation of a Cooper pair costs energy.  Upon approaching  the temperature-tuned superconducting transition this energy becomes small, and $\tau_{sc}$ grows as ${\rm ln}^{-1}(T/T_{c0})$. Consequently, the AL contribution to the longitudinal conductivity~\cite{aslam} is:
	\begin{equation}
	\delta\sigma_{xx}^{AL} = \frac{e^2}{16\hbar}\ln^{-1}\left(\frac{T}{T_c}\right).
	\label{eq:al}
	\end{equation}
In amorphous films with moderate disorder, the interaction between quasiparticles and superconducting fluctuations leads to an additional singular contribution to the conductivity. Similar to the AL term, this contribution, known as the Maki-Thompson (MT) term,~\cite{varlamov,maki1968,thomp1970}  diverges as ${\rm ln}^{-1}(T/T_{c0})$.  \text{H}owever, the MT term depends also on the dephasing time $\tau_{\varphi}$:
	\begin{equation}
	\delta\sigma_{xx}^{MT} = \frac{e^2}{8 \hbar}\ln\left(\frac{\ln{T/T_c}}{\hbar/k_BT\tau_{\varphi}}\right)\ln^{-1}\left(\frac{T}{T_c}\right).
	\end{equation}
This contribution  is expected to be less significant in inhomogeneous systems.~\cite{charkap} Note that the two expressions given above correspond to films in which the superconducting fluctuations are essentially two-dimensional (2D), while the quasiparticles are three dimensional.

\begin{figure}
\centering
\includegraphics[width=\columnwidth]{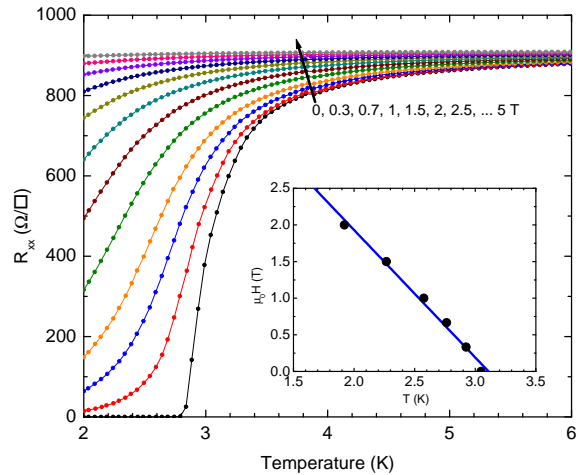}
\caption{ \footnotesize \setlength{\baselineskip}{0.8\baselineskip} Resistive transitions in applied perpendicular magnetic fields between 0 T and 5 T for sample 2; circles are the experimental data, and lines are a guide to the eye. Inset: Upper critical field \text{H}$_{c2}(T)$ versus temperature; here $\mu_0$\text{H}$_{c2}(T)$ was extracted at the point where the resistance approaches 50\% of its normal-state value. The slope $d\text{H}_{c2}/dT \sim$ 1.7 T/K is extracted from a linear fit to the data, and the mean-field transition temperature $T_{c0} \approx$ 2.8 K for this sample.}
\label{fig:lowfield}
\end{figure}

In Fig.~\ref{fig:resistive} we present the zero-field superconducting transition of sample 1; the normal state sheet resistance at 10 K is $R^n_{xx}$ = 0.94 k$\Omega / \Box$. Also shown in Fig.~\ref{fig:resistive} is the resistance measured in an applied field of 8 T, well above $\mu_0$$\text{\text{H}}_{c2}(0) \sim$ 5 T.  The TaN$_x$ film studied in this work can be treated as two-dimensional with respect to superconducting fluctuations.  Dense measurements of the resistive transition as a function of temperature and applied perpendicular magnetic field near $T_c$ on sample 2, shown in Fig.~\ref{fig:lowfield}, were used to extract $\mu_0$$d\text{H}_{c2}/dT\approx$ 1.7 T/K near $T_c$.  The superconducting coherence length $\xi(0) \approx 8.4$ nm is larger than the  film thickness, $d=$4.9 nm. \text{H}all measurements indicate a carrier density of $n\approx 9.1\times 10^{22}$ cm$^{-3}$ for both samples, from which we find that the bulk penetration depth is $\sim 20$ nm. In the presence of disorder the penetration depth increases to $\lambda \sim 140$ nm, while for a 2D film the relevant magnetic screening length becomes $\lambda_\bot =\lambda^2/2d \sim 1970$ nm. The mean free path is estimated to be $\ell \approx$ 0.2 nm. Measured and calculated film parameters for both samples are summarized in Table ~\ref{tab:devparams}.

\begin{table*}[bt]
\begin{center}
\caption{\label{tab:devparams} Measured and calculated TaN$_x$ film experimental parameters. The normal state sheet resistance R$_{xx}$ and carrier density n are measured at 10 K. The transition temperature $T_{c0}$ is extracted from analysis of the fluctuation conductivity. The slope of the upper critical field d\text{H}$_{c2}$/d$T$ evaluated at $T_{c0}$ is extracted from analysis of the resistive transitions of sample 2 in an applied magnetic field. The Ginzburg-Landau coherence length $\xi(0)$, London penetration depth $\lambda_L$, in-plane penetration depth $\lambda_{\bot}$, Ginzburg-Landau parameter $\kappa$, and diffusion coefficient D, are calculated using dirty-limit expressions.~\cite{hsukap}}
\begin{ruledtabular}
\begin{tabular}{c c c c c c c c c c c}
Sample					& $d$ &	R$_{xx}$ &	n	&	$T_{c0}$ & $\mu_0$ d\text{H}$_{c2}$/dT & $\xi$(0) &	$\lambda_L$ & $\lambda_{\bot}$ &	D & $\kappa$ \\
	&	(nm)	& (k$\Omega/\Box$)	&	(cm$^{-3}$)	&	(K)	&	(T/K)	&	(nm)	&	(nm)	&	(nm)	&	(cm$^2$/s) &	\\
\hline
1								&	4.9 &	0.955	&	9$\times$10$^{22}$	&	$\sim$ 2.75		& (1.7) &	8.4 			&	18					&	1980		&	0.51	&	100	\\
2								&	4.9 &	0.944	&	9$\times$10$^{22}$	&	$\sim$ 2.8		& 1.7 	&	8.5 			&	18					&	1970		&	0.51	&	99	\\
\end{tabular}
\end{ruledtabular}
\end{center}
\end{table*}

Adding the AL correction given in Eq.~(\ref{eq:al}) to the normal state resistance (described in the Appendix), we fitted the zero-field resistance as a function of temperature. In the main panel of Fig.~\ref{fig:resistive}, we show that away from to $T_c$ the AL contribution dominates as expected for this class of dirty superconducting films. As the transition is approached, however, the AL expression no longer fits the data. The divergence of the conductivity is stronger than expected from Eq.~(\ref{eq:al}), suggesting that the system is inhomogeneous.  The inset of Fig.~\ref{fig:resistive} shows that close to $T_c$ the conductivity diverges as $\sigma_{xx}\sim(\ln{T}/T_c)^{-1.33}$  which corresponds to pure AL contributions on a percolating cluster.\cite{charkap}  (Note that we do not expect the presence of any such inhomogeneity effects to influence  the \text{H}all conductivity, as was previously shown by Landauer \cite{landauer} from geometrical considerations and by Shimshoni and Auerbach \cite{shimshoni} when quantum effects are included.)  An additional explanation for the departure from the AL term close to $T_{c0}$ can be attributed to the onset of critical fluctuations. This is because the reduced temperature $\ln{T}/{T_{c0}}$ is comparable to the Ginzburg-Levanyuk number\cite{varlamov} Gi for this film:
	\begin{equation}
	G_i = \frac{2 \pi k_B T_{c0} \kappa^2}{\mu_0 \text{H}_{c2}(0) \phi_0 d} \sim 0.06
	\end{equation}
where $\kappa$ is the Ginzburg-Landau parameter.  

Since in this film the superconducting transition is broadened, unambiguous determination of the transition temperature $T_{c0}$ is difficult. This large uncertainty draws into question any quantitative analysis that relies on a precise value for $T_{c0}$. For example, the inset of Fig.~\ref{fig:resistive} shows the calculated fluctuation conductivity $\delta\sigma_{xx}=\sigma_{xx}-\sigma_{n}$  for three different choices of $T_{c0}$. Although all three $T_{c0}$ values fall in the  range of temperatures in which the resistivity drops toward zero, the behavior of $\delta\sigma_{xx}$ as the temperature approaches $T_{c0}$ are different in each case.  Many studies use fits to AL theory to extract $T_{c0}$,~\cite{kokubo01,hsukap} but the AL fits with $T_{c0} = 2.7-2.8$~K shown in Fig.~\ref{fig:resistive} are inconsistent with the fluctuation conductivity in our films, and we can only roughly determine that $T_{c0}$ is about 2.75 K for this sample.  Our analysis of the field dependent \text{H}all effect that follows is acutely sensitive to $T_{c0}$, and this approach may be a particularly useful probe of this parameter.

\begin{figure}[!t]
\centering
\includegraphics[width=1.0\columnwidth]{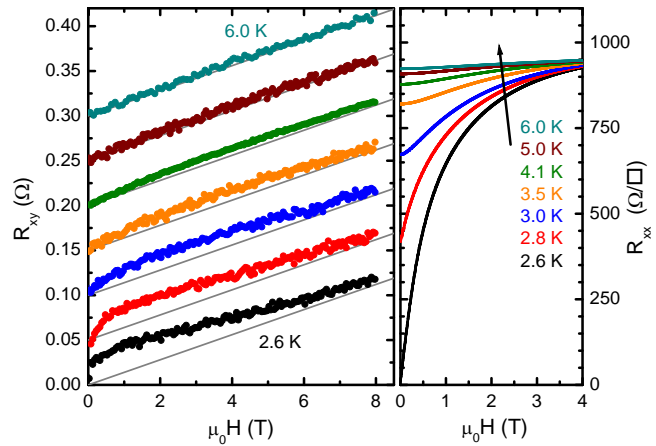}
\caption{ \footnotesize \setlength{\baselineskip}{0.8\baselineskip} Hall resistance $R_{xy}$ (left) and longitudinal resistance $R_{xx}$ (right) of TaN$_x$ sample 1 versus applied magnetic field \text{H} at temperatures near and above the mean-field $T_{c0} \sim 2.75$ K. The \text{H}all resistance curves have been vertically offset for clarity. At temperatures $T \gg T_c$ the \text{H}all resistance is only weakly temperature dependent and is linear in the applied magnetic field, with a slope of $\approx$ 0.014 $\Omega /$ T corresponding to a 3D carrier density of $\sim 9\times 10^{22}$ cm$^{-3}$; this is shown in the thick gray lines.}
\label{fig:rvsh1}
\end{figure}

\section{Hall Effect}

We turn now to the transverse resistance measurements.  Figure~\ref{fig:rvsh1} shows the longitudinal resistance and \text{H}all resistance of sample 1 as a function of applied perpendicular magnetic field, at temperatures close to and above $T_{c0}$.  In the normal phase of a homogeneously disordered film like TaN$_x$, the \text{H}all conductivity is determined by the fluctuations of the order parameter in addition to the quasiparticles, because vortex  physics is not relevant. Thus, it is reasonable to expect that at temperatures above $T_{c0}$ the deviation of the \text{H}all conductivity from the normal state linear-magnetic-field dependence can be attributed to the fluctuations of the order parameter alone.

Fluctuation contributions to the \text{H}all conductivity have been studied using a number of different formalisms. From phenomenological considerations, Lobb \textit{et al.}\cite{lobb94} parameterize the \text{H}all conductivity at temperatures near $T_c$ with terms proportional to the magnetic field $\text{H}$ and $\text{H}^{-1}$
	\begin{equation}
	\sigma_{xy}(\text{H}) =  \frac{c_1}{\text{H}} + c_2 \text{H},
	\end{equation}
which is intended to interpolate between the low-field region where $\sigma \sim \text{H}^{-1}$, and high fields where $\sigma \sim \text{H}$. We find that this simple form does not account for the \text{H}all conductivity seen in TaN$_x$ above $T_{c0}$.

Recent theoretical studies of  superconducting fluctuation contributions to the \text{H}all effect~\cite{michaeli12,tikhonov12} have extended previous calculations of the \text{H}all conductivity~\cite{fukuyama71,aronov95} to a broader range of temperatures and magnetic fields. Close to the critical temperature, Ref.~\onlinecite{michaeli12} shows that the fluctuation \text{H}all conductivity $\delta\sigma_{xy}$ for a broad range of magnetic fields is
\begin{equation}\label{eq:HallCond}
\delta\sigma_{xy} = \frac{2e^2k_BT\varsigma}{\pi \hbar} \text{sign}(\text{H})\sum_{N=0}^{\infty}\frac{(N+1)(\mathcal{E}_{N+1}-\mathcal{E}_{N})^3}{\mathcal{E}_{N}\mathcal{E}_{N+1}(\mathcal{E}_{N}+\mathcal{E}_{N+1})^2}{\textstyle\vert }_{\omega=0}.
\end{equation}
The function $\mathcal{E}_{N}$ describes the superconducting fluctuations in the diffusive regime:
\begin{align}\nonumber
\mathcal{E}_{N}(\omega,\text{H},T) =\ln\left(\frac{T}{T_{c0}}\right) &+ \Psi\left(\frac{1}{2} + \frac{-i\omega+\Omega_c(N+1/2)}{4 \pi k_BT}\right) \\
&-\Psi\left(\frac{1}{2}\right)+\varsigma\omega. \label{ah}
\end{align}
The spectrum of these collective modes is determined by the equation $\mathcal{E}_{N}(\omega,\text{\text{H}},T)=0$. \text{H}ere, $\Psi$ is the digamma function, and $\Omega_c=4|e|\mu_0\text{H}D$ is the energy of the cyclotron motion corresponding to the collective modes (where $D$ is the diffusion coefficient and $k_B$ is Boltzmann's constant).  Since the superconducting fluctuations carry charge, the magnetic field quantizes their spectrum. This  is reflected in the sum over the index $N$ appearing in Eq.~(\ref{eq:HallCond}). The parameter $\varsigma=-\frac{1}{2}\partial{\ln{T_c}}/\partial\mu\propto 1/\gamma \varepsilon_F$ (where $\varepsilon_F$ is the Fermi energy  and $\gamma$ the dimensionless coupling constant of the attractive electron-electron interaction that induces superconductivity) describes the particle-hole asymmetry of the superconducting fluctuations.~\cite{aronov95,michaeli12} For a film with three-dimensional electrons and a simple electron spectrum $\varsigma$ is negative. This parameter, which is essential for the \text{H}all effect, is nonzero due to the energy dependence of the quasiparticle density of states.

The expression given in Eq.~(\ref{eq:HallCond}) corresponds to the AL contribution to the Hall conductivity in the region of classical fluctuations, meaning that it is valid as long as $\mathcal{E}_{0}(\omega=0,\text{\text{H}},T)\lesssim1$. The contributions of the MT kind~\cite{fukuyama71} to the \text{H}all conductivity are less singular and can be disregarded as the transition is approached.  In the limit $\text{H}\rightarrow0$, our result coincides with the one found in Ref.~\onlinecite{aronov95}:
\begin{align}\label{ALH}
\delta\sigma_{xy}=\frac{e^2\varsigma\Omega_c\sign \text{H}}{96\ln^2{T}/{T_{c0}}}.
\end{align}

\begin{figure}[!t]
\centering
\includegraphics[width=0.9\columnwidth]{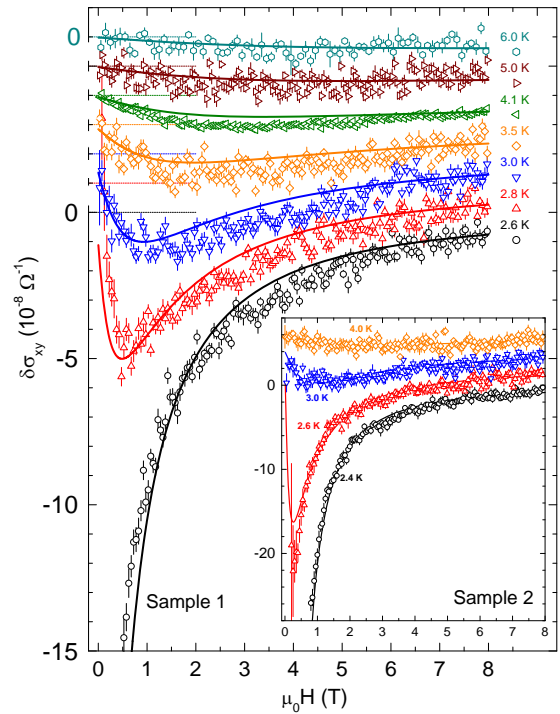}
\caption{ \footnotesize \setlength{\baselineskip}{1.0\baselineskip} Fluctuation Hall conductivity for TaN$_x$ sample 1 at temperatures near $T_{c0}$; the data are offset vertically for clarity. The continuous curves corresponding to Eq.~(\ref{eq:HallCond}) show excellent agreement over a wide range of magnetic fields and temperatures. The inset shows similar data and fits for sample 2.}
\label{fig:flcond}
\end{figure}

To fit the experimental data collected at temperatures close to $T_c$ with Eq.~(\ref{eq:HallCond}), we calculated the \text{H}all conductivity $\sigma_{xy} = - R_{xy}/\left(R_{xx}^2 + R_{xy}^2 \right)$.  Far from the transition, at high magnetic fields and/or temperatures, the measured  \text{H}all resistance is linear in the applied magnetic field and weakly temperature dependent.  This behavior is expected at temperatures  $T\gg T_c$ and/or high magnetic fields where the superconducting fluctuations are insignificant.  Subtracting the normal (linear in magnetic field) component of the \text{H}all conductivity, $\sigma_{xy}^n$, leaves only the fluctuation contribution $\delta\sigma_{xy}$ that is sensitive to the onset of superconductivity
	\begin{equation}
	\delta\sigma_{xy}=\sigma_{xy} - \sigma_{xy}^n.
	\end{equation}
In the temperature range of interest the  \text{H}all resistance measured at $\mu_0 \text{H} = 14$ T changes by $\sim 1\%$; the analysis that follows is insensitive to this slight temperature dependence. We determine the normal state $\sigma_{xy}^n(B)=\sigma_{xy}$(B = 14 \text{T})/(14 \text{T}).  (The analysis that follows is not sensitive to the exact description of the longitudinal normal state resistance, described in the Appendix.)  The fluctuation \text{H}all conductivity $\delta\sigma_{xy}$ calculated for sample 1 is shown in Fig.~\ref{fig:flcond}, along with the fits to  Eq.~(\ref{eq:HallCond}); the inset shows similar data and best-fit curves for sample 2.  The theoretical calculations are in good agreement with the data over a wide range of temperatures and fields.

In fitting Eq.~(\ref{eq:HallCond}) for each sample, three parameters are used for the entire set of Hall conductivity curves: $T_{c0}^{fit}$, D (the diffusion coefficient that enters $\Omega_c$), and  $k_B\varsigma$. For sample 1, the best-fit parameters are (i) $T_{c0}^{fit}=2.60 \pm .05$ K, comparable to that determined from the $R_{xx}$ versus $T$ analysis ($T_{c0} \sim 2.75$ K), (ii) the diffusion coefficient $D=0.52$ cm$^2$/sec, and (iii) the parameter $k_B\varsigma = -3.4\times10^{-4}$ K$^{-1}$.  Note that with a zero-temperature coherence length extracted from $\text{H}_{c2}$, we expect $D\sim \xi^22k_BT_{c0}/\hbar \approx 0.5$ cm$^2$/sec. Estimating $\varsigma$ from the carrier density and taking $\gamma \approx 0.2$ which is suitable for this material, we obtain $k_B\varsigma \sim -10^{-4}$ K$^{-1}$,  in agreement with the values obtained from the fit.  As is evident from Fig.~\ref{fig:flcond}, at $T = 2.6$ K $ \approx T_{c0}$ the magnitude of {\bf $\delta\sigma_{xy}$} sharply increases as the magnetic field decreases and the transition into the superconducting state approaches.  As shown in the inset to Fig.~\ref{fig:flcond} the calculated fluctuation conductivity and fits to Eq.~(\ref{eq:HallCond}) for sample 2, are almost identical to sample 1.  Best-fit parameter values for both samples are listed in Table~\ref{tab:fitparams}.

The above fitting procedure is acutely sensitive to $T_{c0}$ due to the stronger divergence of $\delta\sigma_{xy}$ as the transition is approached [compare Eqs.~(\ref{eq:al}) and~(\ref{ALH})], and provides a precise and clear route to extracting $T_{c0}$.

\begin{table}[bt]
\caption{\label{tab:fitparams}The best fit parameters, $T_{c0}^{fit}$, $D$, and $k_B\varsigma$, for both samples.}
\begin{ruledtabular}
\begin{tabular}{c c c c}
Sample					& $T_{c0}^{fit}$	&	$D$						&	$k_B\varsigma$		\\
								&	(K)							& (cm$^{2}$/s)	&	(K$^{-1}$)				\\
\hline
1								&	2.60$\pm$.05		&	0.52					&	3.4$\times10^{-4}$	\\
2								&	2.53$\pm$.05		&	0.59					&	4.6$\times10^{-4}$	\\
\end{tabular}
\end{ruledtabular}
\end{table}

\section{Conclusion}

In summary, we have performed careful studies of the longitudinal and \text{H}all conductivities at temperatures near and above the zero-field superconducting transition in disordered films of TaN$_x$. Studying fluctuation effects in the \text{H}all conductivity is an experimental challenge in systems with high carrier concentration and large longitudinal resistance.  These measurements appear to be consistent with theoretical analysis over a wide range of temperatures and magnetic fields.  Observation and verification of this effect may facilitate more careful studies of superconducting contributions to the \text{H}all effect and provide a direct route to extract information about the particle-hole asymmetry of the superconducting fluctuations through the parameter {\bf $\varsigma$}.  Finally, such an analysis provides a more precise technique for estimation of the temperature where the gap closes.

We would like to thank the staff of the Stanford Nanocharacterization Laboratory. This work was supported by National Science Foundation grants NSF-DMR-9508419 (AK), NSF-DMR-1006752 (AMF) and by a NSF Graduate Research Fellowship (NPB), as well as Department of Energy Grant DE-AC02-76SF00515. AMF and KM were supported by the U.S.-Israel BSF, and KST by N\text{H}RAP grant.

\appendix

\section*{Appendix: Estimation of the normal state resistance}

\begin{figure}
\centering
\includegraphics[width=\columnwidth]{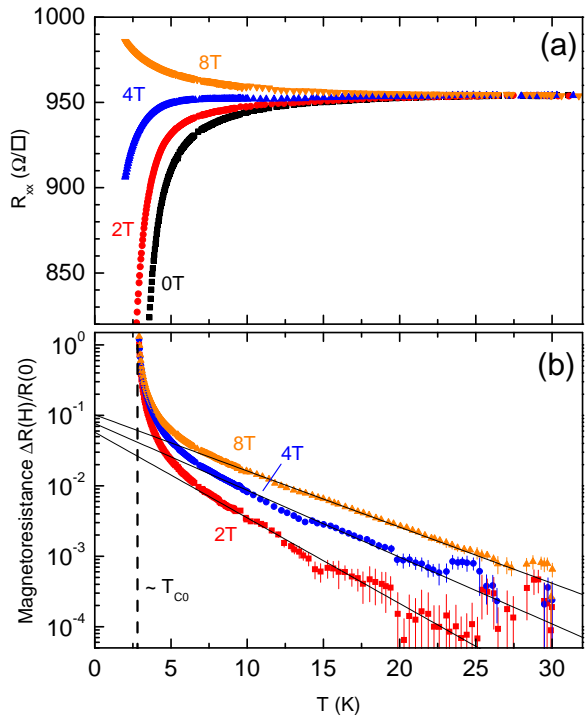}
\caption{ \footnotesize \setlength{\baselineskip}{0.8\baselineskip} Magnetoresistance versus temperature for TaN$_x$ sample 1. Panel (a) shows the measured resistance in applied magnetic fields of 0, 2, 4, and 8 T.  Panel (b) shows the calculated magnetoresistance $\Delta R / R$ as described in the text; the continuous curves are guides to the eye, and the vertical bar indicates the approximate position of $T_{c0}$.}
\label{fig:magres}
\end{figure}

In many studies the normal state resistance is either weakly temperature dependent~\cite{kokubo01} or determined experimentally by applying a large magnetic field to suppress superconductivity and assuming a negligible normal-state magnetoresistance (MR).~\cite{hsukap}   In our samples, a large nonclassical MR prohibits such an approach, and so we need a well defined procedure to account for it.  Figure~\ref{fig:magres} shows the resistance versus temperature of sample 1 up to 30 K for various values of applied magnetic field.  The lower panel of this figure shows the MR, defined as
	\begin{equation}
	\frac{\Delta R}{R}(T,\text{H}) = \frac{R_{xx}(T,\text{H}) - R_{xx}(T,0)}{R_{xx}(T,0)},
	\end{equation}
calcualted using these same data.  According to Kohler's rule\cite{ziman},  the classical normal state MR should be a universal function of $\omega_c\tau$, and in the low-field limit the MR $\Delta R/{R}\sim(\omega_c\tau)^2$, where $\omega_c=e B/m$ is the  cyclotron frequency of the electrons, $\tau$ is their elastic scattering time, B = $\mu_0 \text{H}$, and $\mu_0$ is the magnetic permeability. For our samples $(\omega_c\tau)^2 < 10^{-6}$, much smaller than the measured MR shown in Fig.~\ref{fig:magres}, and the low temperature MR does not scale as a universal function of $\omega_c\tau$.  While we expect superconductivity effects to give a large MR close to the superconducting transition, this behavior should decay to zero as the temperature or magnetic field are increased.  The measured MR at 30 K, well above $T_c \sim$ 2.8 K, is still three orders of magnitude larger than $(\omega_c\tau)^2$, and thus we must describe this large non-classical MR.  Our first step in identifing the normal state resistance $R^{n}$ is to fit the data at 10 K $<$ T $<$ 30 K and various magnetic fields with the following function: 
\begin{equation}\label{MR}
	\frac{\Delta R}{R}(T,\text{H}) = A(\text{H}) \rm{exp}\left[ - T / T_0(\text{H}) \right].
\end{equation}
By interpolation we can find the phenomenological parameters A(\text{H}) and T$_0$(\text{H}) for arbitrary fields, and hence, obtain an expression for the normal state MR at all T and \text{H}.  To now determine the normal state resistance at zero field, R$^{n}$(T,\text{H}=0), we use the expression for the normal state MR given in Eq.~(\ref{MR}) and the resistance measured at 8 T
\begin{align}
&R^{n}_{xx}(T,\text{H}=0) =\frac{R_{xx}(T,\mu_0\text{H}=8\text{T})}{\frac{\Delta R}{R}(T,\mu_0\text{H}=8\text{T})+1}\\\nonumber
&= \frac{R_{xx}(T,\mu_0 \text{H}=8\text{T})}{A(\mu_0 \text{H}=8\text{T}) \rm{exp}\left[ - T / T_0(\mu_0 \text{H}=8\text{T}) \right]+1}.
\end{align}
This approach necessarily recovers the measured zero-field resistance at high temperatures, assuming that deviations from the exponential temperature dependence of the MR at low temperature arise from superconducting fluctuations. Now we are fully equipped to estimate the normal-state resistance for any temperature and magnetic field $R^{n}_{xx}(T,\text{H})$ both close and far away from the transition
\begin{equation}
R^{n}_{xx}(T,\text{H}) = \frac{R^{n}_{xx}(T,0)}{A(\text{\text{H}}) \rm{exp}\left[ - T / T_0(\text{\text{H}}) \right] + 1}.
\end{equation}
Recent studies~\cite{rullier2011} of fluctuation phenomenon in high-$T_c$ materials have used a similar approach to describe normal state behavior.

\end{document}